\documentclass[final,1p,times]{elsarticle}

\usepackage{amssymb}

\journal{Nuclear Physics A}

\begin{document}

\begin{frontmatter}



\title{The nature of $\Lambda(1405)$ hyperon resonance in chiral dynamics}


\author[YITP]{D. Jido},

\author[Kyoto]{T. Sekihara},

\author[Tokyo,RIKEN]{Y. Ikeda},

\author[TITECH]{T. Hyodo},

\author[YITP]{Y. Kanada-En'yo},

\author[IFIC]{E. Oset}

\address[YITP]{Yukawa Institute for Theoretical Physics, Kyoto University, Kyoto 606-8502, Japan}

\address[Kyoto]{Department of Physics, Graduate School of Science, Kyoto University, Kyoto, 606-8502, Japan}

\address[Tokyo]{Depertment of Physics, University of Tokyo, Tokyo 113-0033, Japan}

\address[RIKEN]{Nishina Center for Accelerator-Based Science, Institute for Physical
and Cemical Research (RIKEN), Wako, Saitama 351-0198, Japan}

\address[TITECH]{Department of Physics, Tokyo Institute of Technology, Meguro 152-8551, Japan }
	
\address[IFIC]{Departamento de F\'{\i}sica Te\'orica and IFIC,
Centro Mixto Universidad de Valencia-CSIC,
Institutos de Investigaci\'on de Paterna, Aptdo. 22085, 46071 Valencia, Spain}



\begin{abstract}
The nature of the $\Lambda(1405)$ is discussed based on the unitarised 
coupled-channels approach with chiral dynamics (chiral unitary model). 
This approach describes the $\bar KN$ scattering cross sections 
and the $\Lambda(1405)$ spectra phenomenologically very well.
With this successful description of $\Lambda(1405)$, it is found that 
the $\Lambda(1405)$ is composed by two resonance states having 
different coupling nature to the meson-baryon states. As a consequence,
the resonance 
position in the $\pi\Sigma$ invariant mass spectrum depends on the 
initial channel of the $\Lambda(1405)$ production.
To observe the $\Lambda(1405)$ initiated by the $\bar KN$ channel,
$K^{-} d \to \Lambda(1405) n$ is one of the most 
favorable reactions. 
%
%
Hadronic molecule states with kaons are also discussed by 
emphasizing an important role of $\Lambda(1405)$ as a 
quasibound state of $\bar KN$

\end{abstract}


\end{frontmatter}


\section{Introduction}

The nature of the $\Lambda(1405)$ resonance is an important
issue particularly to understand $\bar K$-nucleus interactions.
Being located just below the threshold of $\bar KN$, 
the $\Lambda(1405)$ is a key resonance for the study of subthreshold kaons in nuclei. 
Due to the decay of the $\Lambda(1405)$ to $\pi \Sigma$, 
for the structure of the $\Lambda(1405)$, both 
$\pi \Sigma$ and $\bar KN$ dynamics are essential.
The $\Lambda(1405)$ has been a historical example of 
a dynamically generated resonance in 
meson-baryon coupled-channels dynamics with $S=-1$~\cite{Dalitz:1959dn}.
Modern treatments based on chiral dynamics with a unitary framework, 
reproduce well the observed spectrum of the $\Lambda(1405)$ together 
with cross sections of $K^{-}p$ to various 
channels~\cite{Kaiser:1995eg,Oset:1998it,Oller:2000fj,Oset:2001cn,Hyodo:2002pk,Jido:2003cb,Borasoy:2005ie,Borasoy}.
This method is based on the chiral perturbation theory
of an effective theory of QCD at low energies giving 
fundamental interaction of meson and baryon, 
and scattering theory, such as the $N/D$ method, to guarantee unitarity 
on the scattering amplitudes. Analyticity of the scattering amplitude 
is also essentially important to obtain 
subthreshold amplitudes and to investigate resonance properties 
in the complex energy plane. 
The obtained amplitude reproduces the meson-baryon scattering well 
and provides $\Lambda(1405)$
as a dynamically generated resonance in $s$-wave meson-baryon scattering. 

On the success of the theoretical description of the $\Lambda(1405)$, 
it it worth noting the following facts; 
1) the available experimental data to constrain theoretical descriptions of 
the scattering amplitude with $S=-1$ around the $\Lambda(1405)$ energies
are the $\bar KN$ scattering and kaonic hydrogen data. 
Thus, the present theoretical description lacks 
experimental information of $\pi \Sigma \to \pi \Sigma $, which may be important to study 
$\bar K$ in few-body nuclear systems. 
2) the famous $\Lambda(1405)$ spectrum given in Ref.~\cite{Hemingway:1984pz}
is a $\pi^{-}\Sigma^{+}$ invariant mass observed 
in the $\Sigma^{+}(1660) \to \Lambda(1405) \pi^{+}$ decay. Therefore,
the initial channel to produce $\Lambda(1405)$ is not clear. 
As seen below, since the $\Lambda(1405)$ spectrum shape depends on the 
production channel, it is extremely important to control the initial channel 
experimentally and to investigate the production mechanism theoretically 
for the good understanding of the nature of $\Lambda(1405)$.


\section{The structure of $\Lambda(1405)$}
In this section, we explain the findings about the nature of $\Lambda(1405)$ obtained 
by the chiral unitary approach. 

\subsection{Double pole nature of $\Lambda(1405)$}

Resonances are expressed as poles of the scattering amplitude in the complex 
energy plane. 
In the chiral unitary approach for the meson-baryon system developed 
by Ref.~\cite{Oller:2000fj,Oset:2001cn}, the scattering amplitude is obtained 
in an analytic form, so that one can easily perform analytic continuation of the
scattering amplitude to the complex energy plane.
In Ref.~\cite{Jido:2003cb} it was found that 
the $\Lambda(1405)$ seen in the $\pi \Sigma$ invariant mass spectra 
consists of two resonance poles located at the $\Lambda(1405)$ energies, 
and these two state have different coupling properties to the meson-baryon states;
one is located at $1390 - 66 i$ MeV (pole 1) with a larger width having strong coupling to the 
$\pi \Sigma$ state, ant the other pole is at higher energy $1426 - 16i$ (pole 2) MeV with 
a narrower width and dominantly couples to the $\bar KN$ state. 
Because these two states closely appear with the large widths, one cannot observe 
these states separately in the $\pi \Sigma$ invariant mass spectra (See Fig.~\ref{fig:MSI0}(b)). 
What we can see is a single peak structure provided by the interference 
of these two states, that is the $\Lambda(1405)$ spectrum observed in experiments.

The reason that there exist two poles around the $\Lambda(1405)$ energies 
is that there are two attractive channels with $S=-1$ and $I=0$ 
in $s$-wave at these energies~\cite{Jido:2003cb},  $\bar KN$ and $\pi \Sigma$. 
These two attractive channels provide the two resonance states. 
It is also found in a recent work~\cite{Hyodo:2007jq} that 
the $\Lambda(1405)$ can be described essentially by two coupled channels of 
$\bar KN$ and $\pi\Sigma$ rather than a complete SU(3) configuration by 
four channels, $\bar KN$, $\pi\Sigma$, $\eta \Lambda$ and $K \Xi$ with $I=0$. 
In addition, without the channel couplings between $\bar KN$ and $\pi\Sigma$, 
one has a bound state in the $\bar KN$ channel below the $\bar K N$ threshold and 
a  resonance in the $\pi \Sigma$ channel above the $\pi \Sigma$ threshold.
These two states appear very close to the original states 
obtained in coupled channels. Therefore, the essential ingredients of $\Lambda(1405)$ 
is the $\bar KN$ bound state and $\pi \Sigma$ resonant correlation.  
This is the reason for the double pole structure of $\Lambda(1405)$ 
and the different coupling nature of the two states. 

\subsection{Channel dependence of $\Lambda(1405)$ spectrum}

\begin{figure}
\centerline{\includegraphics[width=13.5cm]{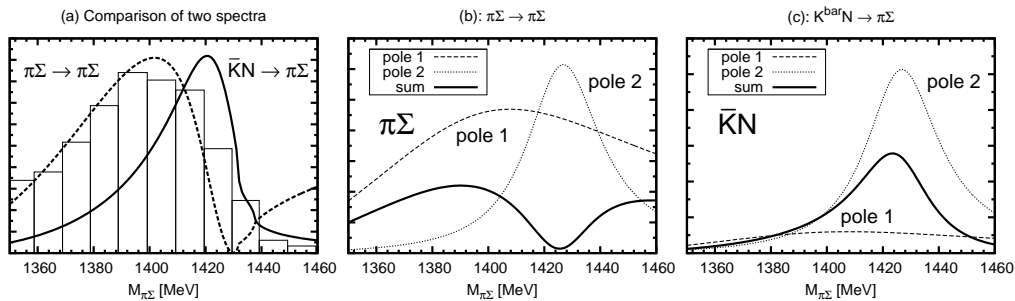}}
\caption{$\pi\Sigma$ invariant mass spectra with $I=0$ in arbitrary units~\cite{Jido:2003cb}. 
(a): chiral unitary model calculation of the $\pi \Sigma$ invariant mass spectra of  
$\bar KN \rightarrow \pi\Sigma$ (solid line) and 
$\pi\Sigma \rightarrow \pi \Sigma$ (dashed line).
The histogram denotes an experimental data given in Ref.~\cite{Hemingway:1984pz}.
(b) and (c): simple model calculations of the $\pi\Sigma$ invariant mass spectra initiated 
by $\pi\Sigma$ and $\bar KN$, respectively, in which the scattering amplitudes are given by 
two Breit-Wigner terms.
The dashed (dotted) line denotes the spectrum only with pole 1 (pole 2) term, 
while the solid line shows the spectrum calculated by coherent sum of pole 1 and 2. 
The Breit-Wigner parameters are 
determined by the chiral unitary model. 
For the details, see Ref.~\cite{Jido:2003cb}. 
\label{fig:MSI0}}
\end{figure}

The presence of the two poles having different coupling properties
results in initial channel dependence of the $\Lambda(1405)$ peak 
position in the $\pi\Sigma$ invariant spectrum.
In Fig.~\ref{fig:MSI0}(a), we show the $\pi \Sigma$ invariant 
mass spectra with $I=0$ initiated by two different channels, $\bar KN$ and 
$\pi \Sigma$~\cite{Jido:2003cb}.
The figure shows that the $\Lambda(1405)$ peak  is seen in 
the different energies;
For the $\pi \Sigma$ initiated spectrum, the resonance peak appears at around 1405 MeV,
whereas in the $\bar KN \to \pi\Sigma$ channel, the peak position is located higher 
at around 1420 MeV instead of 1405 MeV. 
This is because, for the $\pi\Sigma$ channel, both of the lower and higher poles 
contribute to the spectrum, while the $\bar KN$ system selectively couples to the higher 
pole, as seen in Figs.~\ref{fig:MSI0}(b) and~\ref{fig:MSI0}(c), where we show
decomposition of the $\Lambda(1405)$ spectrum into each pole contribution.
The decomposition is performed by the model calculation in which
the amplitudes are given by Breit-Wigner terms for pole 1 and 2 together with 
the resonance parameters, such as mass, width and couplings, determined by the 
chiral unitary model~\cite{Jido:2003cb}. 
In this way, the resonance position in the $\pi \Sigma$ invariant mass
depends on the channels to produce the $\Lambda(1405)$. 
Particularly, the finding that the $\Lambda(1405)$ resonance appears at 1420 MeV 
in the $\bar KN$ initiated channel is important for kaon-nucleus systems,  
since the resonance appearing in the $\bar KN$ channel 
is the one relevant for the kaon-nucleon interaction.

One of the experimental confirmation of the double pole nature of the $\Lambda(1405)$
is to compare the $\Lambda(1405)$ spectra in different channels.
Especially, as we already discussed above, one of the important 
consequences is that  the resonance 
in the spectrum initiated by the $\bar KN$ channel appears at 1420 MeV.
Therefore, it is interesting to see the resonance position
in the $\Lambda(1405)$ production initiated by $\bar KN$. 
However,  direct production $\bar KN \to \Lambda(1405)$ 
is kinematically forbidden, since the $\Lambda(1405)$ resonance appears 
below the $\bar KN$ threshold. This fact leads us to indirect 
productions of $\Lambda(1405)$.
But, in the indirect production, it is necessary to investigate
the $\Lambda(1405)$ production mechanism theoretically 
to recognize the initial channels of the $\Lambda(1405)$ production.

\subsection{$\Lambda(1405)$ observed in $\bar KN$ channel}

\begin{figure}
\centerline{\includegraphics[width=5cm]{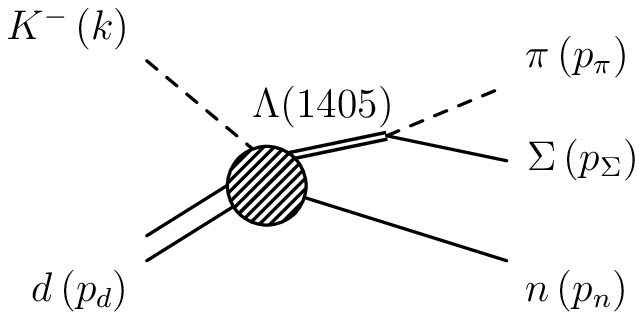}
\hspace{0.5cm} \includegraphics[width=7.5cm]{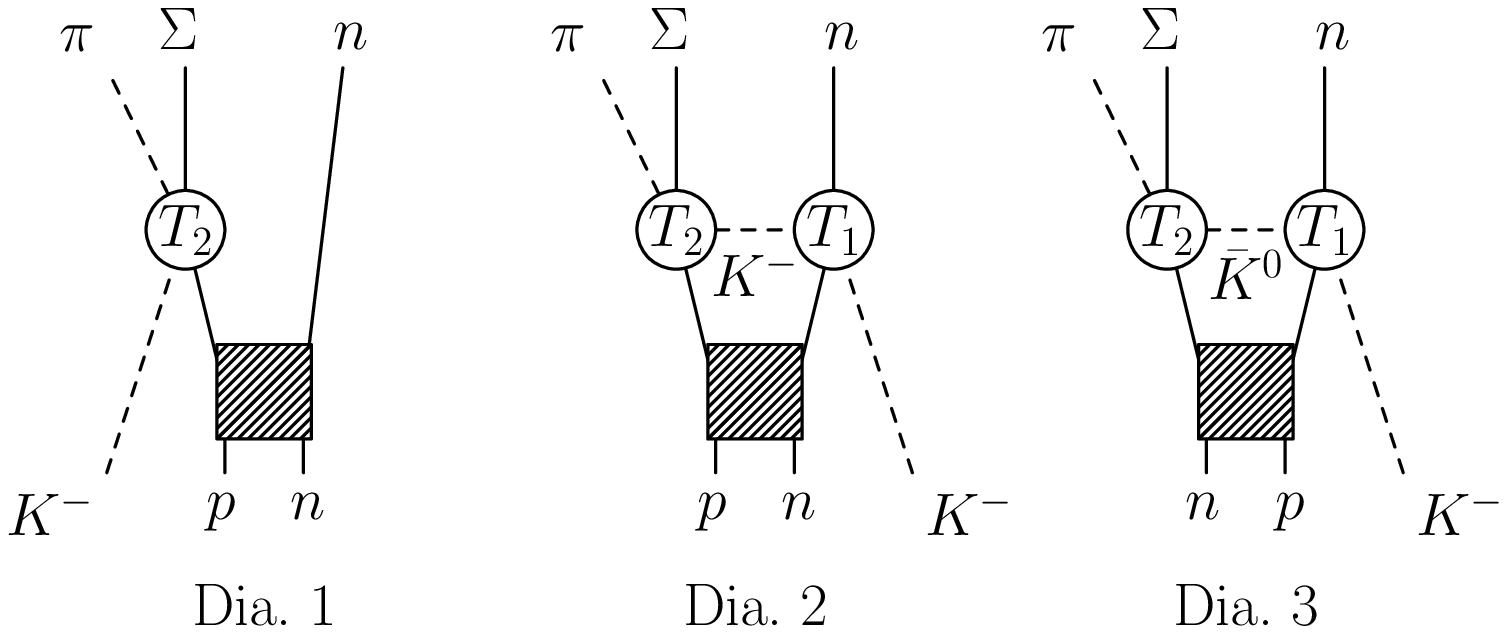}}
\caption{Diagrams for the calculation of the $K^{-}d \to \pi\Sigma n$ reaction.
The right three are the Feynman diagrams for our calculation. 
$T_{1}$ and $T_{2}$ denote the scattering amplitudes for $\bar KN \to \bar KN$
and $\bar K N \to \pi \Sigma$, respectively.  \label{fig2}}
\end{figure}

Here we discuss the $K^{-}$ induced production of $\Lambda(1405)$
with a deuteron target, $K^{-} d \to \Lambda(1405) n$, proposed in Ref.~~\cite{Jido:2009jf},
to see the $\Lambda(1405)$
in the $\bar KN$ channel. In this reaction, the final neutron takes energy out from the 
initial kaon.  
The produced $\Lambda(1405)$ decays into $\pi \Sigma$ with $I=0$
as shown in Fig.~\ref{fig2}, and the $\Lambda(1405)$ is identified by the 
$\pi \Sigma$ invariant mass spectra. 
The relevant contribution for the $\Lambda(1405)$ production in this reaction is given
by three diagrams  shown in Fig.~\ref{fig2}. 
The amplitude $T_{1}$ and $T_{2}$ denote $s$-wave scattering amplitudes of 
$\bar K N \to \bar KN$ and $\bar K N \to \pi \Sigma$, respectively, and 
the $\Lambda(1405)$ resonance is involved in the amplitude $T_{2}$.
Thus, it is important noting that, in this reaction, 
the $\Lambda(1405)$ is produced selectively by the $\bar KN$ channel.  
This is because the strangeness is brought into the system by the initial kaon.

Diagram 1 of Fig.~\ref{fig2} expresses  the $\Lambda(1405)$ production 
in the impulse approximation. Diagrams 2 and 3 are for two-step
processes with $\bar K$ exchange.
For the energetic incident $K^{-}$ with several hundreds MeV/c momentum
in the lab.\ frame, the contribution of diagram 1 (direct production) is 
expected to be very small, since the $\Lambda(1405)$ is produced 
by a far off-shell nucleon and the deuteron wavefunction has tiny component of such 
a nucleon. In contrast to the direct production, in the double
scattering diagrams, the large energy of the incident $K^{-}$ is carried 
away by the final neutron and the exchanged kaon can have a suitable
energy to create the $\Lambda(1405)$ colliding with the other nucleon
in the deuteron. 
Double scattering diagrams with pion exchanges hardly contribute to the 
$\Lambda(1405)$ production, since $\Sigma$ and $\pi$ are emitted separately
from the $T_{1}$ and $T_{2}$ amplitudes.  
Such diagrams may give 
smooth backgrounds in the $\pi \Sigma$ invariant mass spectra.

\begin{figure}[bt]
\centerline{\includegraphics[width=5.5cm]{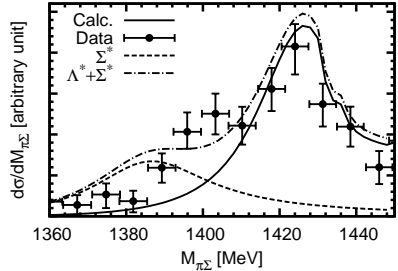}}
\caption{
$\pi \Sigma$ invariant mass spectra of $K^{-}d \rightarrow \pi^{+}\Sigma^{-}n$ in arbitrary units
at 800 MeV/c incident $K^{-}$ momentum.
The solid line denotes the present calculation.
The data are taken from the bubble chamber experiment at $K^{-}$ momenta between 686
and 844 MeV/c given in Ref.~\cite{Braun:1977wd}.
A possible inference of the $\Sigma(1385)$ resonance to the 
$\Lambda(1405)$ spectrum of $K^{-}d \rightarrow \pi^{+}\Sigma^{-}n$ 
at 800 MeV/c incident $K^{-}$ momentum. 
The $\Sigma(1385)$ spectrum (dashed line) is calculated 
by the Breit-Wigner amplitude.
The dash-dotted lines denotes an incoherent sum of the $\Lambda(1405)$ 
and $\Sigma(1385)$ spectrum. 
\label{fig:IMspecSig}}
\end{figure}

We show, in Fig.~\ref{fig:IMspecSig},
the $\pi^{+}\Sigma^{-}$ invariant-mass spectrum 
in arbitrary units at 800 MeV/c incident $K^{-}$ momentum and compare 
our theoretical calculation with the experimental data.  
The data are taken from  the bubble chamber experiment at $K^{-}$ momenta between 
686 and 844 MeV/c~\cite{Braun:1977wd}. 
In the analysis of the experiment, the resonance contribution was determined by fitting 
a relativistic Breit-Wigner distributions and a smooth background 
parametrized as a sum of Legendre polynomials to the data. 
We show,  in Fig.~\ref{fig:IMspecSig}, the resonance (foreground) contributions for 
the experimental data which are obtained by  
subtracting the background contributions from the actual data points given in 
the paper.
The solid line denotes our theoretical calculation with the scattering amplitudes
$T_{1}$ and $T_{2}$ obtained by chiral unitary approach. 
The spectrum shape obtained in this calculation agrees with that of the experimental observation.
Especially it is very interesting to see that the peak position, 
which comes from the $\Lambda(1405)$ production, 
appears around $M_{\pi\Sigma}=1420$ MeV instead of 1405 MeV. 
This is one of the strongest evidences that the resonance position 
of the $\Lambda(1405)$ depends on the initial channel of meson and baryon,
and supports the double pole nature of the $\Lambda(1405)$,
in which the higher state sitting in 1420 MeV strongly couples to the $\bar KN$ channel. 

The bump structure seen around $M_{\pi\Sigma}=1400$~MeV in data is 
related to the $p$-wave contributions coming from the $\Sigma^{*}(1385)$ 
production~\cite{Jido:2009jf,Yamagata}, 
which were not taken into account in the present calculation. 
In Ref.~\cite{Jido:2009jf}, an estimation of a possible 
influence of the $\Sigma(1385)$ resonance on 
the $\Lambda(1405)$ spectrum appearing around 1420~MeV
was discussed
by calculating the $\Sigma(1385)$ spectrum in the Breit-Wigner 
formulation with the mass 1385 MeV and the width 37~MeV
and the phase space factor $|\vec p_{\pi}^{\, *}|\,|\vec p_{n}|$.
Summing up the spectra of the $\Lambda(1405)$ and $\Sigma(1385)$ 
incoherently, they find that the peak structure at 1420~MeV
is not affected by the $\Sigma(1385)$ contribution as seen in 
Fig.~\ref{fig:IMspecSig} (dash-dotted line). 
In this estimation, they have adjusted the height
of the $\Sigma(1385)$ spectrum so as to reproduce the observed 
bump structure around 1390 MeV. 
A further quantitative calculation of the $\Sigma(1385)$ spectrum
by taking into account of  the $\Sigma(1385)$ production 
mechanism in the present reaction 
with the $p$-wave contributions following Ref.~\cite{Jido:2002zk},
will be coming soon~\cite{Yamagata}.

\subsection{Model dependence of pole positions and significance of
 $\pi\Sigma$ scattering data}

The higher pole located at around 1420 MeV is essentially a $\bar KN$ quasibound 
state and strongly couples to $\bar KN$. It has been pointed out 
in Ref.~\cite{Hyodo:2007jq} that the position of the higher pole has
less model dependence among different calculations based on coupled-channels 
approach with chiral dynamics, while these calculations predict 
different positions for the lower pole which couples to $\pi\Sigma$ strongly
(see Fig.~8 in Ref.~\cite{Hyodo:2007jq}). This is because, in these model calculations,
$K^{-}p$ scattering data are used to fix the model parameters and the properties 
of the $\bar KN$ interaction are well constrained by exciting data.
It was also pointed out in a different approach~\cite{Ikeda:2007nz} that 
a pole appears at around 1420 MeV in the $\bar KN$ scattering amplitude 
which reproduces the $\bar{K} N$ scattering lengths.
Contrarily,
we have few data available for $\pi\Sigma \to \pi\Sigma$ scattering so far.

The lack of the $\pi\Sigma$ scattering data makes the properties of 
the lower pole less controlled in theoretical calculations. 
Any information of the $\pi \Sigma$ scattering is desired to fix further 
the nature of the $\Lambda(1405)$.
For instance, the threshold properties of the $\pi\Sigma$ scattering, such
as the scattering length and effective range, could give suitable constraint~\cite{IHJKSY}.
In Table~\ref{tab:pisigma}, we show values of the $\pi\Sigma$ scattering length with $I=0$  
calculated using various chiral coupled-channels models. The values of the scattering 
length are substantially dependent on the models, which predict also different pole 
positions of the lower pole.  

\begin{table}
\begin{center}
\begin{tabular}{|c|cccc|c|}
\hline
model & BNW~\cite{Borasoy:2005ie}  & ORB~\cite{Oset:2001cn} & HNJH~\cite{Hyodo:2002pk} & BMN~\cite{Borasoy} & virtual state \\
\hline
$a_{\pi\Sigma}$ [fm] & 0.517 & 0.789 & 0.692 & 0.770  &  $\sim5$ \\
$z_{1}$ [MeV]& $1388-39i$ & $1389-64i$ & $1400-76i$ & $1440-76i$ &1325 (V) \\
\hline
\end{tabular}
\caption{Model calculations of the $I=0$ $\pi\Sigma$ scattering lengths $a_{\pi\Sigma}$ 
using various chiral coupled-channels models. 
(For the detail, see Ref.~\cite{IHJKSY}). 
$z_{1}$ denotes the pole position of the lower resonance. 
The right row shows the value of the $\pi\Sigma$ scattering length calculated 
under the assumption that there is a  virtual state around 
10 MeV below the $\pi \Sigma$ threshold. \label{tab:pisigma}}
\end{center}
\end{table}

A more interesting question is whether the pole in the $\pi \Sigma$ is really a resonance 
or not. The chiral unitary approach predicts that the $\pi \Sigma$ strong 
correlation forms a resonance state found above the $\pi\Sigma$ threshold. 
Yet we do not have the experimental confirmation of the $\pi\Sigma$ resonance pole. 
If the $\pi \Sigma$ interaction were unexpectedly strong enough, there could exist 
a $\pi \Sigma$ virtual state below the $\pi \Sigma$ threshold in stead of the resonance 
state. In this case, the $\Lambda(1405)$ would be consisted of one single pole. 
Alternatively, for the $s$-wave interaction, energy-independent potentials can provide 
only virtual states. Thus, in case that the energy-dependence of the $\pi\Sigma$ interaction 
would be weak in the $\pi\Sigma$ threshold energies, the $\pi \Sigma$ attraction 
could provide a virtual state. 
This sort of meson-baryon amplitude is studied
in Ref.~\cite{Ikeda:2007nz}.
%
The question of either resonance or virtual state
in the $\pi\Sigma$ correlation can be answered by the magnitude of the $\pi \Sigma$
scattering length. If the virtual state exists close to the threshold, the scattering length 
is in order of 5 fm, while, as we already saw, for the resonance case, the scattering length 
is around 0.5 fm, which is one order of magnitude small.

Understanding the $\pi\Sigma\to\pi\Sigma$ scattering properties is important 
not only for the $\Lambda(1405)$ structure but also for $\bar K$ in few-body systems,
such as $\bar K NN$~\cite{akaishi02,shevchenko07,Ikeda:2007nz,dote09}. 
As discussed in Ref.~\cite{Jido:2008zz} for  
the $K \bar KN$ system, if $\bar K$ few-body states are weakly bound, 
the $\bar K$ and nucleons are the essential constituents, and $\pi$ and $\Sigma$ 
can be regarded as nonactive degrees of freedom. Thus, the coupled channels
effect of $\pi \Sigma$ can be  incorporated into the effective $\bar KN$ interaction
as done in Ref.~\cite{Hyodo:2007jq}, and the few-body bound state is essentially 
obtained in the single channel problem. However, since the $\pi\Sigma$ threshold 
is located 100 MeV below the threshold of the $\bar KN$, if the $\bar K$ few-body 
systems turn out to be bound with a large binding energy,  inclusion of the $\pi\Sigma$ 
coupled channel effect to theoretical calculations are unavoidable~\cite{Ikeda:2008ub},
and there in-vacuum $\pi\Sigma\to\pi\Sigma$ scattering data are necessary 
as a fundamental input.

\subsection{$\Lambda(1405)$ as meson-baryon quasibound state}

The third consequence of the chiral unitary model is that $\Lambda(1405)$
is almost purely a quasibound state of the meson and baryon, and quark components
are much less important. A recent analysis~\cite{Hyodo:2008xr} have shown that, 
although in the chiral unitary approach only the dynamics of meson and baryon 
is considered, the formulation implicitly involves 
some other components than the meson and baryon considered in the model
space of the framework. The work~\cite{Hyodo:2008xr} also discussed how to exclude the
implicit components in theoretical description of scattering amplitudes, and compared 
the pole positions of the $\Lambda(1405)$ in the theoretical amplitude with those of 
the scattering amplitude obtained phenomenologically so as to reproduce  the
observed $\bar KN$ scattering. They find that these two descriptions provide
very similar pole positions of $\Lambda(1405)$ and 
conclude that the $\Lambda(1405)$ is predominantly
described by the meson-baryon component.

Owing to the meson-baryon quasibound structure, the $\Lambda(1405)$ has a different
nature than typical baryon resonances which may have largely three-quark components. 
For instance, as a theoretical analysis, 
the behavior with the number of colors ($N_c$) of the 
$\Lambda(1405)$ has been discussed in Ref.~\cite{Hyodo:2007np}, and it 
has been found that the $N_c$ behavior of the decay widths is  different 
from the general counting rule for a $qqq$ state.
Another example is the size of the $\Lambda(1405)$.
In Ref.~\cite{Sekihara:2008qk},
the electromagnetic mean squared radii of the $\Lambda(1405)$ were calculated.
They found that the electric radius of  the $\Lambda(1405)$ is about three times 
larger than that of the neutron. 
This is a consequence of the small binding energy ($10\sim30$ MeV), in which
the constituent kaon in the $\Lambda(1405)$ is an almost real particle surrounding around
the nucleon.

\section{$\Lambda(1405)$ in few-body systems}

The $\Lambda(1405)$ can be an building-block of $\bar K$ nuclear
few-body systems, since one of the pole in the $\Lambda(1405)$ is a quasibound
state of $\bar KN$. It is quite interesting that the $\Lambda(1405)$ can be reproduced 
by nonrelativisitic potential models for the $\bar KN$ system~\cite{akaishi02,Hyodo:2007jq}.
This is because the binding energy ($10\sim30$ MeV) is not so large in comparison
with typical hadron energy scale. (In other words, the kaon kinetic energy in 
the $\bar KN$ bound system is much smaller than the kaon mass). 
Thus, if the few-body systems also have not so large binding energies,
the single channel potential model is one of the reasonable approached 
for the kaonic few-body systems, as done in Refs.~\cite{akaishi02,dote09} for $\bar KNN$.
This picture will be broken down if the binding energy is so large that 
the other channels become active. In such a case, the coupled channels effects
should be properly treated in theoretical calculations, as done 
in Refs.~\cite{shevchenko07,Ikeda:2007nz} for $\bar KNN$.

Here we would like to present other kaonic few-body systems,
$K \bar K N$ and $\bar K \bar K N$ systems with $I=1/2$ and $J^{P}=1/2^{+}$.
In Ref.~\cite{Jido:2008zz}, these systems were investigated in
a nonrelativistic three-body potential model
under the assumption that the $\Lambda(1405)$ resonance and the scalar 
mesons, $f_{0}(980)$, $a_{0}(980)$, are reproduced as 
quasibound states of $\bar KN$ and $K \bar K$, respectively. 
The effective two-body interactions are described by complex-valued functions
representing the open channels, 
($\pi \Lambda$, $\pi \Sigma$) for $\bar K N$ and
($\pi \pi$, $\pi\eta$) for $K\bar K$. 
They found a quasibound state
of the $K \bar K N$ system around 1910 MeV 
below all of the meson-baryon decay threshold energies of 
the $\Lambda(1405)+K$, $f_0(980)+N$ and 
$a_0(980)+N$ states,
which means that the obtained bound state is stable against breaking up to
the subsystems. 
This quasibound state was also confirmed later by a more sophisticated calculation 
using a relativistic Faddeev approach~\cite{MartinezTorres:2007sr}.
For the $\bar K \bar K N$ system, the binding energy from 
the $\Lambda(1405)+\bar K$ threshold was found to be as small as a few MeV
due to the strong repulsion $\bar K \bar K$ with 
$I=1$~\cite{Jido:2008zz}.

For the structure of the $K\bar KN$ system,
it was found  that  the $\bar KN$ and $K\bar K$ subsystems 
are dominated by  $I=0$ and $I=1$, respectively,
and that these subsystems have very similar 
properties with those in the isolated two-particle systems. 
This leads to the picture that the $K \bar KN$ system
can be interpreted as coexistence state of $\Lambda(1405)$
and $a_{0}(980)$ clusters, and $\bar K$ is a constituent
of both $\Lambda(1405)$
and $a_{0}(980)$ at the same time, as seen in the $\bar KNN$ 
sysytem~\cite{yamazaki07,dote09}.
Consequently, the binding energy and width of 
the $K\bar KN$ state 
is almost the sum of those in $\Lambda(1405)$ and $a_{0}(980)$.
It is also found that the inter-hadron distances in the $K\bar KN$ state
are larger than 2 fm, which is comparable to typical 
nucleon-nucleon distances in nuclei. Therefore, 
the $K\bar KN$ system more spatially extends 
than typical baryons described by quark models.  
These features are caused by weak binding of the three hadrons,
for which the $KN$ repulsion plays an essential role.

The finding that the $\Lambda(1405)$ keeps its properties 
in few-body systems motivates that 
picture that the $\Lambda(1405)$ resonance can be a doorway state of $\bar K$
absorption to nuclear systems~\cite{Sekihara:2009yk}.
Especially the coupling strengths of the $\Lambda(1405)$ to $\bar KN$ and $\pi\Sigma$
are important parameters to understand the non-mesonic transitions 
$\Lambda(1405)N \to YN$ ($Y=\Lambda$ or $\Sigma$), which may be 
dominant processes of the nonmesonic $\bar K$ absorption in nuclei.

\section{Conclusion}

The $\Lambda(1405)$ is the ``gift'' of the meson-baryon dynamics. 
Due to the strong attractions in the $\bar KN$ and $\pi \Sigma$ channels,
there exists two resonance poles around the $\Lambda(1405)$ energies.
One is essentially a $\bar KN$ quasibound state just below the $\bar KN$
threshold, the other is a $\pi \Sigma$ resonance having a wide width and
strong couping to the $\pi \Sigma$ state.
The interference of these two states forms a single 
resonance peak in the $\pi \Sigma$ invariant mass spectra.
Therefore the $\pi \Sigma$ invariant mass spectra depend on the initial
channel for the $\Lambda(1405)$ production. 
The $\Lambda(1405)$ produced by the $\bar KN$ channel can
be observed by $K^{-} d \to \Lambda(1405)n$. For better understanding
of the $\Lambda(1405)$ structure and $\bar K$ nuclear few-body 
systems, any experimental information of the $\pi \Sigma \to \pi \Sigma$ 
scattering are desired, especially the $\pi\Sigma$ scattering length
can fix the lower pole position of $\Lambda(1405)$. 
$K\bar KN$ can be another example of the $\bar K$ nuclear few-body 
systems. It is found that the $K \bar KN$ quasi-bound system
is interpreted as coexistence state of $\Lambda(1405)$
and $a_{0}(980)$ clusters and $\bar K$ is a constituent
of both $\Lambda(1405)$ and $a_{0}(980)$ at the same time.



\section*{Acknowledgments}
This work is supported in part by
the Grant for Scientific Research (No.~20028004) from Japan Society for 
from the Ministry of Education, Culture,
Sports, Science and Technology (MEXT) of Japan.
A part of this work is done under Yukawa International Project for 
Quark-Hadron Sciences (YIPQS).

\end{document}